\documentclass[a4paper,11pt]{article}
\pdfoutput=1
\usepackage{jheppub}
\usepackage[T1]{fontenc}

\def\ep{\varepsilon}
\def\half{\frac{1}{2}}
\def\com#1#2{\left[#1,#2\right]}
\def\K{\mathcal{K}}
\def\lr#1{\left(#1\right)}
\def\slr#1{\left[#1\right]}

\def\trl#1{\textrm{Tr}\lr{#1}}

\def\CPn{\mathbb C P^n}
\def\CPFn{\mathbb C P^n}
\def\RS2{\mathbb R\times S^2_F}

\def \be  {\begin{equation}}
\def \ee  {\end{equation}}
\def \bex  {\begin{equation*}}
\def \eex  {\end{equation*}}
\def \bea {\begin{eqnarray}}
\def \eea {\end{eqnarray}}
\def \bal {\begin{align}}
\def \eal {\end{align}}
\def\quater{\frac{1}{4}}
\def\tFo{\left._2 F\right._1}
\def\no{\nonumber\\}
\def\labell#1{\label{#1}}
\def\bibitemm#1{\bibitem{#1}}

\def \PRD {{Phys. Rev. D\ }}
\def \JHEP {{JHEP\ }}
\def \RMP {{Rev. Mod. Phys.\ }}

\allowdisplaybreaks

\title{\boldmath Uniform order phase and phase diagram of scalar field theory on fuzzy $\mathbb C P^n$ }

\author[a]{Juraj Tekel}
\affiliation[a]{Department of Theoretical Physics\\Faculty of Mathematics, Physics and Informatics, Comenius University\\Mlynska Dolina, 842 48 Bratislava, Slovakia}
\emailAdd{tekel@fmph.uniba.sk}

\abstract{
We study the phase structure of the scalar field theory on fuzzy $\mathbb C P ^n$ in the large $N$ limit. Considering the theory as a hermitian matrix model we compute the perturbative expansion of the kinetic term effective action under the assumption of distributions being close to the semicircle. We show that this model admits also a uniform order phase, corresponding to the asymmetric one-cut distribution, and we find the phase boundary. We compute a non-perturbative approximation to the effective action which enables us to identify the disorder and the non-uniform order phases and the phase transition between them. We locate the triple point of the theory and find an agreement with previous numerical studies for the case of the fuzzy sphere.
}

\begin{document} 
\maketitle
\flushbottom

\section{Introduction}
Fuzzy spaces are finite mode approximation to compact manifolds \cite{fuzzy2,fuzzy}. Since their algebra of functions is finite dimensional, they provide a way to regularize quantum field theories without breaking the symmetry of the underlying space. Fuzzy spaces are a part of a broader framework of noncommutative geometry \cite{cones1,noncom2}, with applications in many areas of theoretical physics, from condensed matter \cite{cpn} to string theory \cite{string}. Field theories on fuzzy spaces are subject of intense research and among many other aspects, the phase structure of the scalar field theories has been studied both numerically and analytically, see \cite{panero} for a review.

It has been found that there are three different phases in the theory. A disorder phase where the field oscillates around the zero, a uniform order phase where the field oscillates around a non-zero value and a non-uniform order, or striped, phase where the field does not oscillate around one fixed value. This third phase was shown to be absent in the commutative field theories \cite{XXX}. Which of the three phases is realized depends on the parameters of the theory and the studies have also identified three phase transition lines which all meet at one point. This structure creates a phase diagram of the theory.

%Its use spreads many areas of the theoretical physics, from condensed matter \cite{XXX} to particle physics \cite{XXX} and string theroy \cite{XXX}.
The original idea behind the introduction of the noncommutative framework to the field theory, which was the hope that the short distance structure of the spacetime will help to cure the loop divergences \cite{snyder}, turned out to be spoiled by a phenomenon of UV/IR mixing \cite{uvir2}. This is the most prominent feature of the noncommutative theories and will play an important role also in the presented work. Its root is in the non-locality of the noncommutative theory causing an interplay of short and long distance processes and makes the theory non-renormalizable, even in the commutative limit.

Other manifestation of the UV/IR mixing can be found in the described phase diagram of the fuzzy field theories. The striped phase is not present in the phase diagrams of the commutative field theories and its presence in the large $N$ limit of the fuzzy phase diagram points to the difference between the commutative theory and the commutative limit of the noncommutative theory.

We will use matrix models techniques to analyze the phase structure of the fuzzy fields theories. Since the field is a hermitian matrix, the field theory can be viewed as a particular matrix model with a new term in the probability distribution corresponding to the kinetic term of the action. The phase structure of the field theory is then determined by the eigenvalue distribution of the matrix. This was first considered in \cite{steinacker2}. In \cite{ocon,samann} the effect of the kinetic term was studied perturbatively and the disorder to non-uniform order phase transition was identified. Non-pertubative treatment of the kinetic term was later considered in \cite{our1,our2}. More recently, it was shown how to fully incorporate the kinetic term of the action into the matrix model distribution in the case of the fuzzy sphere, for symmetric distributions that are not too different from the semicircle distribution, \cite{poly}. We will give more details about this effective action method in section \ref{seceffaction}.

The presented work consists of two main parts. In the first, we will generalize the effective action procedure to the case of of fuzzy $\CPFn$ and obtain a perturbative expansion of the effective action. We will find a non-perturbative approximation to this action, which will enable us to compute the phase boundary between the disorder and the non-uniform order phases, i.e. the one-cut/two-cut phase transition of the matrix model. In the second part, we will introduce minimal asymmetry into the effective action and show, that this modification yields a model which does admit also an asymmetric one cut distribution, representing the uniform order phase. We will find the region in parameter space where this configuration is allowed and find the triple point of the theory, which is the point where the two phase boundaries meet. Several numerical studies have located this triple point in the case of the fuzzy sphere. Our result will turn out to be reasonably close to these results, especially considering rather bold approximations we have made.

\section{Preliminaries and review of some previous results}
In this section we will introduce some basic notions needed for the rest of the paper. We will discuss the formulation of fuzzy field theory and some related matrix model techniques we shall use. We will be brief and mention only the most necessary information. For more details, we direct the reader to the referenced reviews and original work.

\subsection{Matrix models and phase transitions}

In a hermitian random matrix model \cite{saclay}, the object of interest is a random hermitian $N\times N$ matrix with a probability distribution $P(M)=e^{-S(M)}$ and the expected value of a matrix function $f(M)$ given by
\be
\left\langle f\right\rangle=\frac{1}{Z}\int dM e^{-S(M)}f(M)\ ,\labell{matav}
\ee
where $Z$ is a normalization factor. The integration measure is the Haar measure on the group of all hermitian matrices. As long as both $S(M)$ and $f(M)$ are invariant under $M\to U M U^\dagger$, we can diagonalize the matrix $M$ and we can treat the problem as a problem of $N$ eigenvalues $x_i$ of the matrix $M$. The Jacobian factor turns out to be the Vandermonde determinant
\be
dM=dU\Bigg(\prod_i dx_i\Bigg)\Bigg(\prod_{i<j}(x_i-x_j)^2\Bigg)\ ,
\ee
and can be incorporated into the action as an effective repulsive potential. We obtain
\be
\left\langle f\right\rangle=\frac{1}{\tilde Z}\int \prod_{i=1}^N dx_i\, f(x_i)e^{-\slr{S(x_i)-2\sum_{i<j}\log|x_i-x_j|}}\ ,
\ee
where the changed normalization includes the angular integral over the measure on the $SU(N)$ group ${\int}dU$. We are going to be interested only in the large $N$ limit of such expressions. The second term in the exponential is a sum of ${\sim}N^2$ terms and thus of the same order. We need also $S$ to be of the order $N^2$ in this limit to contribute. This means that in the large $N$ limit the integral will be dominated by the stationary configuration of the eigenvalues.

Our prime goal will be the eigenvalue distribution of the random matrix in quartic model, i.e. the model with the probability distribution given by
\be
S(M)=\trl{\half r M^2+gM^4}=\lr{\half r \sum x_i^2+g \sum x_i^4}\ .\labell{mataction}
\ee

For a positive $r$, the famous result for the eigenvalue distribution in the large $N$ limit is the deformed Wigner semicircle distribution \cite{brezin}
\be
\rho(x)=\frac{1}{\pi}\lr{\frac{r}{2}+g a^2+2gx^2}\sqrt{a^2-x^2}\ , \ x^2<a^2\ , \ 	a^2=\frac{1}{6g}\lr{\sqrt{r^2+48g}-r}\ .
\ee
The $g=0$ case is the celebrated Wigner semicircle for the gaussian hermitian ensemble.

For negative values of $r$, this distribution is not always non-negative. For some combinations of $r,g$ it becomes negative, can not be interpreted as a probability distribution and for such values the solution for the eigenvalue distribution is different. It is supported on two intervals and referred to as a two-cut distribution, opposing to the one-cut support of the former case. The boundary line between the two regions is given by
\be
r(g)=-4\sqrt{g}\labell{purepot}
\ee
and when we change the parameters in (\ref{mataction}) such that we cross this line, the eigenvalue distribution enjoys a phase transition from the one-cut to the two-cut phase.  If the parameters of the theory are from the part above the line, the eigenvalue distribution will be of the one-cut type, if the part under the line, the eigenvalue distribution will be of the two-cut type. We will refer to this division of the parameter space as a phase diagram.

\subsection{Scalar fuzzy field theory (as a random matrix model)}

These are field theories defined on the fuzzy spaces \cite{fuzzy2}, of which the prototypical is the fuzzy sphere $S^2_F$. The coordinates on $S^2_F$ are defined by the relations
\be \hat x_i \hat x_i = \rho^2 \ , \ \hat x_i \hat x_j - \hat x_j \hat x_i=i \theta \ep_{ijk} \hat x_k\ ,\ee
which are realized by the $N$ dimensional representation of $SU(2)$
\be \hat x_i=\frac{2r}{\sqrt{N^2-1}} L_i \ , \ \theta=\frac{2r}{\sqrt{N^2-1}} \ , \ \rho^2=\frac{4r^2}{N^2-1} j(j+1)=r^2\ ,\ee
where $L_i,i=1,2,3$ are the generators of $SU(2)$ in this representation. Scalar field on the fuzzy sphere is an arbitrary polynomial in $\hat x_i$, i.e. $N{\times}N$ hermitian matrix. It can be decomposed into the polarization tensors basis $T^l_m$, which are the eigenmatrices of $L^2$, i.e
\be
M=\sum_{l=0}^{N-1}\sum_{m=-l}^l c_{l,m} T^l_m \ , \ L^2 T^l_m\equiv[L_i,[L_i,T^l_m]]=l(l+1)T^l_m\ .\labell{expansion}
\ee
In this setting, the analogues of the derivative and the integral are a commutator with $L_i$ and a trace, so with analogy to commutative theory, we can define the field theory action
\bea
S(M)&=&\trl{-\half [L_i,M][L_i,M]+\half rM^2}+V(M)\no&=&
	\trl{\half M[L_i,[L_i,M]]+\half rM^2}+V(M)\ .
\eea
The overall volume factor has been absorbed into the definition of the matrix and the coupling constants. The field theory in terms of the correlation functions can then be constructed by the matrix version of the Feynman diagrams. We will deal with a generalized case of the $\phi^4$ field theory and the action will be defined by
\be
S=\trl{\half M \K M+\half rM^2+gM^4}\ ,\labell{THEaction}
\ee
where we assume that the kinetic term to be diagonal in the $T^l_m$ basis $\K T^l_m=K(l)T^l_m$ and we further assume that $K(0)=0$. The standard case is given by $K(l)=l(l+1)$.

The field theory is defined by the correlation functions and, by the virtue of the field being a matrix, is equivalent to a matrix model with probability distribution (\ref{THEaction}) and we can use the matrix model techniques to analyze the theory in the large $N$, i.e. commutative, limit.\footnote{As $N\to\infty$ we get $\theta\to0$.} However the diagonalization trick is not going to work anymore, since the kinetic part of the action is not invariant under $M\to U M U^\dagger$ and the angular integral will not be trivial anymore. In \cite{ocon,samann} this integral was explicitly computed perturbatively by group theoretical methods. In \cite{our1}, it was shown that for the free scalar field theory with an arbitrary kinetic term, the large $N$ limit of the eigenvalue distribution of the corresponding matrix is again a Wigner semicircle with a rescaled radius
\be\labell{ourR}
R=2\sqrt{\frac{f(r)}{N}}\ ,
\ee
where
\be
f(r)=\sum_{l=0}^{N-1}\frac{2l+1}{K(l)+r}\labell{fS2}\ .
\ee

As mentioned in the introduction, the field theory exhibits three different phases, separated by the transition lines. The disorder phase corresponds to the one-cut distribution of the matrix model, the non-uniform order phase to the two-cut distribution and the uniform order phase to an asymmetric one-cut distribution. This distribution was not present in the phase diagram of the matrix model (\ref{mataction}) and is a novel feature of the matrix models given by the fuzzy action (\ref{THEaction}).

\subsection{Kinetic term effective action}\labell{seceffaction}

In this section we briefly describe a method to integrate the angular degrees of freedom developed in \cite{poly}. As a result, we will obtain additional terms to the action (\ref{mataction}). We will mention only the main idea, approximations made and the important results. For more details, the reader is referred to the original work.

\paragraph{Effective action.} The angular integral has to be a function of the eigenvalues, and thus a function of the moments
\be m_n=\trl{M^n}=\sum x_i^n\ .\labell{moments} \ee
 With a further assumption that the eigenvalue distribution is symmetric, $m_{2n+1}=0$, and that the identity matrix has a vanishing kinetic term, we obtain
\be
\int dU e^{-\trl{\half M\K M}}=e^{-S_{eff}(t_n)} \ , \ t_n=\trl{M-\frac{1}{N}\trl{M}}^n \ , \ n=2,4,\ldots\ .\labell{polyexpansion}
\ee
Finally, when we write the equation of motion for the eigenvalue distribution, the Wigner semicircle has to be a solution. Knowing the moments $t^W_{2n}$ of this distribution and plugging them into the equation of motion we get a constraint on the form of $S_{eff}$. It turns out it can be brought into form of an arbitrary function of $t_2$ plus some additional terms. These extra terms vanish for the semicircle distribution, when $t_n=t^W_n$, and start at the eight order in the eigenvalues
\be
S_{eff}=\half F(t_2)+(b_1+b_2 t_2)(t_4-2t_2^2)^2+\ldots\ .
\ee
As a first approximation we can neglect these extra terms.

We are thus led to the conclusion, that (for eigenvalue distributions close to the semicircle), the effective action is a function of $t_2$ only. From now on, we will denote $t_2\equiv t$.

To compute $F(t)$, the starting point is the result of \cite{our1}, the theory with an action $S=\trl{\half M\K M +\half z M^2}$ yields a semicircular distribution with radius (\ref{ourR}). This theory is, after the integration, equivalent to the theory with an effective action $S_e=\half F(t)+\half z \trl{M^2}$ and the radii of the distributions for these two theories have to be the same
\be
2\sqrt{\frac{f(z)}{N}}=2\sqrt{\frac{N}{F'(t)+z}}\ .
\ee
Also, we need to make sure that the distribution corresponding to the effective action yields a distribution with the correct second moment $c_2$, so we impose a self-consistency condition
\be
t=\int dx\, x^2 \rho(x)=\frac{N^2}{F'(t)+z}\ .
\ee
As a result, the two conditions that determine the kinetic term effective action are
\begin{subequations}\labell{effact}
\bea
F'(t)+z&=&\frac{N^2}{t}\labell{effact1}\ ,\\
t&=&f(z)\ .\labell{effact2}
\eea
\end{subequations}
These equations have a formal solution of the form
\be
F(t)=N^2\int dt\lr{\frac{1}{t}-g(t)}.\labell{formal}
\ee
Here $g(t)$ is the inverse of the function $f(z)$. The scaling $F\sim N^2$ is determined such that the effective action is of the same order as the potential terms and $t$ is appropriately rescaled such that $g(t)$ is also of order $N^2$.

For the case of the fuzzy sphere, $f(z)$ can be computed explicitly, inverted and the integral calculated, to give the kinetic term effective action
\be
F_{S^2_F}(t)=N^2\log\lr{\frac{t}{1-e^{-t}}}\labell{FS2}.
\ee

\paragraph{Phase transition.}
The effective action changes the behavior of the original matrix model and thus shifts the phase transition line (\ref{purepot}). We keep in mind that that the phase transition happens when the distribution becomes negative and we need to impose a similar self-consistency condition for the second moment. The conditions for the phase transition are then, with the factors of $N$ already scaled out using the scaling found in the next section,
\begin{subequations}
\bea
F'(t)+r&=&-4\sqrt g\ ,\labell{trans1}\\
t&=&\frac{1}{\sqrt g}\ ,\labell{trans2}
\eea
\end{subequations}
which yields
\be
r(g)=-4\sqrt g - F'(1/\sqrt{g})\ .\labell{12bound}
\ee
In \cite{poly} it was show that for the fuzzy sphere case (\ref{FS2}), the corresponding result
\be
r(g)=-5\sqrt g - \frac{1}{1-e^{1/\sqrt g}}\ .\labell{12boundSF2}
\ee
matches the previously obtained numerical results \cite{num09}, as well as the perturbative calculation \cite{samann}. The line (\ref{12boundSF2}) divides the parameter space into two parts.

\section{Symmetric one-cut to two-cut phase transition}\labell{sec3}
In this section we compute the effective action $F(t)$ of the kinetic term for the case of the fuzzy $\CPFn$ using the method described in the previous section. We will be very brief about the construction and properties of the fuzzy projective spaces, for more details see \cite{cpn}. The result is in the form of a power series in powers of $t$, which turns out not to be usable to compute the one-cut/two-cut phase boundary. But we will develop a method to compute this line also.

\subsection{Effective action for $\mathbb CP^n$}\labell{sec31}

It is straightforward to generalize the formula for $f(z)$ for the case of $\CPFn$. The kinetic term is the same as for the sphere $\K M=\com{L_a}{\com{L_a}{M}}$, however now the external matrices $L_a$ are the generators of the $N$ dimensional representation of $SU(n+1),a=1,2,\ldots,n^2-1$. The matrix $M$ again decomposes into the irreducible representations of $SU(n+1)$, labeled by $l$ up to a maximal value $L$, analog of (\ref{expansion}). The relationship between $L$ and $N$ is now more complicated, given by
\be
N=\frac{(n+L)!}{n!L!}\ .
\ee
Note, that there is not a value of $L$ for every value of $N$, so the size of the matrix is not arbitrary. However, we can still take a large $N$ limit by taking $L\to\infty$.

We have $K(l)=l(l+n)$, so (\ref{fS2}) becomes 
\be
f(z)=\sum_{l=0}^L \frac{\textrm{dim}(n,l)}{l(l+n)+z}\ ,
\ee
where $\textrm{dim}(n,l)$ is the dimension of the rank-$l$ representation of $SU(n+1)$. In the large $N$ limit, substituting $l=L\,x$ and using 
\be
\textrm{dim}(n,l)\sim\frac{l^{2n-1}}{(n-1)!^2n} \ , \ N\sim\frac{L^n}{n!}\ , 
\ee
the sum becomes an integral
\be
f(z)=2n N^2 \int_0^1 dx \frac{x^{2n-1}}{L^2 x^2+z}\ .
\ee
As a result, we obtain
\be
f(z)=\frac{N^2}{z}\tFo\lr{1,n;n+1,-\frac{(n!N)^{2/n}}{z}}\,
\ee
where $\tFo$ is the hypergeometric function.

Clearly, we will not be able to solve the conditions (\ref{effact}) analytically. Equations involved in inverting $\tFo$ are transcendental and the inverse cannot be expressed in elementary functions. However, we will be able to find the effective action $F(t)$ up to any given order in $t$ by a perturbative calculation.

We express $z$ from (\ref{effact1}), use it in (\ref{effact2}), expand both $F(t)$ and $f(z)$ in powers of $t$ and require the equation to hold order by order. This results in the following expression
\bea
F(t)=N^2&\Bigg[&\frac{n(n!)^{2/n}}{n+1}\lr{N^{2/n-2}t}-\frac{n(n!)^{4/n}}{2(n+1)^2(2+n)}\lr{N^{2/n-2}t}^2-\no&-&\frac{2(n-1)n(n!)^{6/n}}{3(n+1)^3(6+5n+n^2)}\lr{N^{2/n-2}t}^3
-\ldots\Bigg]\ .\labell{effaction}
\eea
with terms growing in complexity in a rather unilluminating way.\footnote{It is clearly tempting to rescale $t$ to incorporate also the $(n!)^{2/n}$ factors, but as this does not seem to provide much simplification in later results, we will proceed further without this rescaling.} We present the explicit formulas up to order $t^7$ in the appendix \ref{aB}. The important observation is that we can rescale $t\to \tilde t N^{2-2/n}$ to make all the terms of the same order in $N$.

This is a good point to mention the issue of scaling of the parameters of the model. For all the terms in the action to contribute, we need them to scale the same way in the $N\to\infty$ limit and this scaling to be $N^2$. For the effective action part we got the scaling $\tilde tN^{2-2/n}=t $ which means $\tilde M N^{\theta_x}=M $ with $\theta_x=\half-\frac{1}{n}$. Similarly, we need to rescale $r$ and $g$ to fix the scaling of the mass and interaction terms, which yields
\be
\theta_r=\frac{2}{n} \ , \ \theta_g=\frac{4}{n}-1\ .\labell{scaling}
\ee
Note, that these scalings are the same as in \cite{poly} for the $\CPFn$ laplacian, which scales as $\alpha=\frac{2}{n}$. Moreover, this scaling agrees with the scaling for the cases $n=1,2,3$ found in \cite{samann} using perturbative techniques.

As a check of (\ref{effaction}), for the case of the fuzzy sphere $n=1$, we obtain the first terms of the expansion of (\ref{FS2}). Further, we can compare this formula with the results of \cite{samann}, where the appropriate formulas for the cases $n=2,3$ were computed using a different method.\footnote{This work provides also the fuzzy sphere result, but it was shown in \cite{poly} that this result is reproduced by (\ref{FS2}).} We give them here in the original notation
\begin{subequations}\labell{sam}
\bea
\beta S^{(n=2)}&=&\beta \frac{2l^4}{3}(c_2-c_1^2)-\beta^2 \frac{2l^4}{9}(c_2-c_1^2)^2-\labell{sam2}\\&-&\beta^3\frac{8l^4}{405}\slr{8(c_2-c^2_1)^3+5\lr{2c_1-2 c_1 c_2+c_3}^2},\no
\beta S^{(n=3)}&=&\beta \frac{2l^5}{4}(c_2-c_1^2)-\beta^2 \frac{3l^4}{20}(c_2-c_1^2)^2-\labell{sam3}\\&-&\beta^3\frac{l^4}{25}\slr{10(c_2-c^2_1)^3+3\lr{2c_1-2 c_1 c_2+c_3}^2}.\nonumber
\eea
\end{subequations}
To connect these with our result, we need to set the odd moments $c_1$ and $c_3$ to zero, use $t=4 \beta l^n c_2/n!$ and recall that the contribution from the effective action is $\half F(t)$.\footnote{Note, that the factor $l^n /n!$ comes from a definition of $c_n$, which is different from $t_n$ by a factor of $N\sim L^n/n!$.} After the substitutions in (\ref{effaction}), we do indeed recover both (\ref{sam2}) and (\ref{sam3}).

From this point on, we will assume the scaling (\ref{scaling}) and drop the tilde sign. We will also drop any $N$ dependence, since under this scaling the large $N$ limit of the tilded quantities is well behaved.

We also make an important observation, which will be important in the next section. If we take $t=(c_2-c_1^2)$ in (\ref{effaction}) we obtain the first two terms of (\ref{sam}) including the odd term $c_1$. This means that up to the fourth order in eigenvalues, the approximation is exact. We also obtain a part of the third term, but not all of it.\footnote{Note, that the rest of the terms can be expressed as $t_3^2$, in the spirit of expansion (\ref{polyexpansion}) including the odd moments $t_{2n+1}$.} Terms neglected due to assumed symmetry of the distribution thus start to emerge at the sixth order in eigenvalues. This is sooner than the terms that emerge due to the departure from the Wigner semicircle distribution, which start at the eight order in eigenvalues. This also summarizes the approximations we have made, namely that the eigenvalues are distributed in a way that is not too asymmetric and close to the semicircle distribution.

\subsection{One-cut to two-cut phase transition}\labell{32}

At this point, we would like to compute the one-cut/two-cut phase transition line for the case of a general $\CPFn$. Not having an explicit formula for the effective action, we cannot compute $F'(t)$ needed in (\ref{12bound}). We could take the expansion (\ref{effaction}) and simply obtain
\be
r=-4\sqrt g - A_1 - \frac{2A_2 }{\sqrt g}- \frac{3A_3 }{g}- \frac{4A_4 }{g^{3/2}}-\ldots\labell{onetotwo}\ ,
\ee
where $F(t)=\sum A_k t^k$. However examining this series for the case of the fuzzy sphere, where we know all the coefficients from (\ref{12boundSF2}), we see that it does not converge to the true solution for small values of $g$. This can be linked to the non-analyticity of the true phase boundary. We therefor need to come up with a reasonable method how to approximate the effective action from the first few terms of the series (\ref{effaction}).

We will proceed in the following way. From (\ref{effaction}) we see that for small $t$ we expect $F\sim t$ and from (\ref{formal}) we expect $F\sim \log t$ for large values of $t$, thus we write $F(t)=\log\lr{1+t\,h(t)}$ and we will try to approximate the function
\be
h(t)=\frac{1}{t}\lr{e^{F(t)}-1}\labell{ht}
\ee
using the method of Pade approximants \cite{pade}. Namely, we express
\be
h(t)=\frac{a_0+a_1 t^1+\ldots+a_{M_a} t^{M_a}}{1+b_1 t+\ldots+b_{M_b} t^{M_b}}
\ee
and expand both sides of (\ref{ht}) in the powers of $t$. We then chose coefficients $a_n,b_n$ so that the equation holds order by order, up to the order $F(t)$ is known to. Since this order is $t^7$, we set $M_a=M_b=3$. For larger $M$'s, the equation would not fix all the coefficients. 

This procedure yields approximate formulas for the effective action, which are going to be useful later but are rather complicated and seem not to reveal much. We present the explicit formulas for the first few cases in the appendix \ref{aB}, plotted here in the figure \ref{ftcpn}. Note the strange peak and dip in the plots for $n\geq2$. These are most probably only a residue of the approximation method, as they are smaller compared to the approximation with $M_{a,b}=2$.

\begin{figure} [tbp]
\centering % \begin{center}/\end{center} takes some additional vertical space
\includegraphics[width=.75\textwidth]{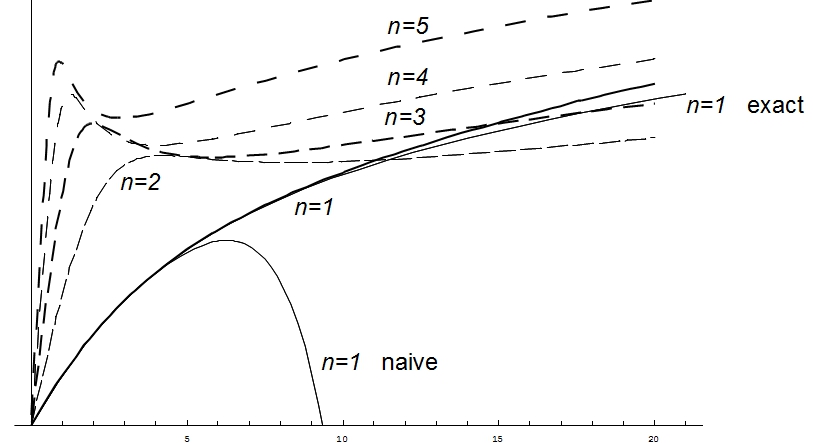}
% "\includegraphics" is very powerful; the graphicx package is already loaded
\caption{The approximation of the effective action $F(t)$ for different cases of $\CPFn$. The exact formula for the fuzzy sphere is (\ref{FS2}) and the naive approximation is given by (\ref{effaction}). Other naive approximations are qualitatively the same and are not plotted.}\labell{ftcpn}
\end{figure}

Finally, we can use the case of the fuzzy sphere as a consistency check. We see that the approximation is pretty good and we expect it to be no worse for other values of $n$.\footnote{For larger values of $t$, the approximate and the exact curves become different. This is due to the special feature of the expansion of $F(t)$ for the fuzzy sphere, where the only non-vanishing odd term is the linear. For the approximation $M_{a,b}=4$, we obtain a correct $\log t$ behavior of the effective action.} We thus conclude, that the effective actions obtained from (\ref{ht}) are a good approximation to the true effective action and we can compute the phase boundary using (\ref{12bound}).

\section{Asymmetric one-cut phase}
We now turn to the most important part of the presented paper, which is the identification of the asymmetric one-cut phase and location of the triple point of the theory. We will start with the observation made at the end of the section \ref{sec31}. If we take $t=(c_2-c^2_1)$ in the expression for the effective action (\ref{effaction}), we will capture some of the asymmetric terms neglected at the beginning and up to second order all of them.

We will thus study the matrix model with the kinetic term effective action
\be
S_{eff}=\half F(c_2 -c_1^2)\approx \half A_1 (c_2 -c_1^2)+ \half A_2 (c_2 -c_1^2)^2\ .\labell{themodel}
\ee
We will show that this model allows for an asymmetric one-cut solution to the eigenvalue distribution. We will analyze the parameter space of the theory and find the two-cut/asymmetric one-cut transition line. We will be able to locate the intersect of this line with the symmetric one-cut/two-cut phase boundary, thus locate the triple point of the theory. We will then compare the computed location with the results of previous numerical simulations for the fuzzy sphere.

We will compute also contribution from the higher order terms in the expression for $F(t)$ and we will study contributions from these terms to the phase boundary and the location of the triple point. The location will not shift significantly and thus we will conclude, that perturbative expansion of the phase boundary for this second phase transition is well behaved in the small $g$ limit and does not require any special treatment. The reason for this is that the phase transition line is analytic at $g=0$ and thus large $g$ expansion does converge.

\subsection{Asymmetric eigenvalue distribution}

We begin this section with a somewhat technical discussion of the derivation of the asymmetric one-cut case of the eigenvalue distribution. The full action of the model which determines the distribution is
\be
S(x_i)=\half A_1 (c_2-c_1^2)+ \half A_2 (c_2-c_1^2)^2+\half r\sum x_i ^2 + g  \sum x_i^4-2 \sum_{i<j}\log|x_i-x_j|\ .
\ee
The saddle point configuration of the eigenvalues is given by the condition
\be
[(A_1+2A_2(c_2-c_1^2)](x_i - c_1)+r x_i + 4 g x_i^3=2\sum_{i\neq j}\frac{1}{|x_i-x_j|}\labell{theeq}\ ,
\ee
where we have used $dc_2/dx_i=2x_i,dc_1/dx_i=1$. In the large $N$ limit, this equation becomes an integral equation for the eigenvalue density $\rho(x)$
\be
-[A_1+2A_2(c_2-c_1^2)]c_1+[r+A_1+2A_2(c_2-c_1^2)] x + 4 g x^3=2P\int dy \frac{\rho(y)}{x-y}\ .\labell{theeqint}
\ee
where $P$ denotes the principal value of the integral. This equation is then solved by the standard matrix models techniques \cite{saclay}, introducing the resolvent
\be
\omega(z)=\int dx\,\frac{\rho(x)}{z-x}\ .
\ee
The eigenvalue distribution can be obtained by a discontinuity equation
\be
\rho(x)=\frac{1}{2\pi i}\Big(\omega(x-i0)-\omega(x+i0)\Big)\ .\labell{rho}
\ee
The coefficients of the $1/z$ expansion of $\omega(z)$ are the moments of the distribution, so we get
\be
\omega(z)=0\cdot z + \frac{1}{z}+\frac{c_1}{z^2}+\frac{c_2}{z^3}+\mathcal O(z^{-4})\labell{seriesresolvent}\ .
\ee
We will use the equation (\ref{theeqint}) to compute the resolvent, leaving $c_1,c_2$ there as parameters. We will then post the selfconsistency conditions, requiring the appropriate terms in the expansion of the solution to match (\ref{seriesresolvent}).

To solve (\ref{theeqint}), we need to make some assumptions about the support. We are after the asymmetric one-cut solution, so we will assume that $\rho(x)$ is nonzero only in the interval $(D-\sqrt \delta,D+\sqrt \delta)$.

This leads to the following four equations, one for each term in (\ref{seriesresolvent}),
\begin{subequations}\labell{mostimportant}
\bea
0&=&-\half c_1 \lr{A_1+2A_2(c_2-c_1^2)}+2 D^3 g + 3 D \delta g +\half D\big(r+A_1+2A_2(c_2-c_1^2)\big)\ ,\\
1&=&3 D^2 \delta g +\frac{3}{4}\delta^2 g \quater\delta\big(r+A_1+2A_2(c_2-c_1^2)\big)\ ,\\
c_1&=&3 D^3 \delta g + \frac{3}{2}D\delta^2 g + \quater D \delta\big(r+A_1+2A_2(c_2-c_1^2)\big)\ ,\\
c_2&=&3 D^4 \delta g + 3 D^2 \delta ^2 g + \quater \delta^3 g + \quater D^2 \delta \big(r+A_1+2A_2(c_2-c_1^2)\big)+\no&+&\frac{1}{16}\delta^2 \big(r+A_1+2A_2(c_2-c_1^2)\big)\ .
\eea
\end{subequations}
It is not difficult to express $c_1,c_2$ in terms of $D$ and $\delta$, but trying to solve the remaining two equations is a lost battle.

However, we can progress further after we realize that we do not need a complete solution. Namely, in the action that led to this system, we took only the linear and quadratic terms in $F(t)$. So any part of the solution that is of a higher than quadratic order is of little interest to us. Technically, we replace $A_1\to \ep A_1$ and $A_2\to \ep^2 A_2$ and solve these equations up to second order in $\ep$.

\paragraph{Zeroth order.} At the zeroth order we recover two interesting solutions. First the symmetric solution
\be
D^{SYM}_0=0\ \ \ ,\ \ \ \delta^{SYM}_0=\frac{-r+\sqrt{r^2+48g}}{6g}\ ,
\ee
which is reassuring, but of little significance at the moment. Second, we get the solution
\be
D_0=\frac{\sqrt{-3 r + 2 \sqrt{r^2-60 g}}}{2\sqrt{5 g}}\ \ \ ,\ \ \ \delta_0=\frac{-r-\sqrt{r^2-60g}}{6g}\ .
\ee
This is the solution we are after, since in the limit $r\to-\infty$, $D_0$ becomes the positive minimum of the true potential $\half r x^2+g x^4$, with a vanishing width $\delta_0$. The solution thus lives in one, but only one of the minima of the potential. We also obtain an identical solution living in the negative minimum of the potential.

\paragraph{First and second order.} To find corrections to this expression, we perturb the zero-th order solution to $\delta=\delta_0+\ep \delta_1+\ep^2 \delta_2$ and $D=D_0+\ep D_1+\ep^2 D_2$, plug them into (\ref{mostimportant}) and solve order by order in $\ep$. As a result, we obtain quite unappealing and untypesetable formulas for $\delta_{1,2},D_{1,2}$ which we happily leave to a computer program to manipulate.

\paragraph{The eigenvalue distribution.} We now have all we need to compute the eigenvalue distribution. Knowing the resolvent as the function of $\delta,D$, and these as a function of $r,g$ and $A_1,A_2$, we obtain from (\ref{rho}) the final formula for the eigenvalue distribution in the asymmetric one-cut case
\bea
\rho(x)&=&\lr{A_1\ep+\half A_2\delta \ep^2+ \frac{1}{8} A_2 \delta^3 \ep^2 g-\frac{9}{8} A_2 D^2 \delta^4 \ep^2 g^2+ 4 D^2 g + 2\delta g +r+4 D g x + 4 g x^2}\no&\times&\frac{1}{2\pi}\sqrt{\delta -(D-x)^2}\ ,\labell{asymdistr}
\eea
where we need to do the $\ep$ expansion, take terms only up to the order $\ep^2$ and then set $\ep=1$.

\subsection{Phase transition and the triple point}

We can now ask the main question. For what values of the parameters does this distribution become negative? Our one-cut assumption is then not valid and this indicates the phase transition. As one easily checks, it happens at the right edge of the interval. Also, we are only after a perturbative solution for the transition line $r(g)$, so we expand $r(g)=r_0+\ep r_1 +\ep^2 r_2$, where $r_{0,1,2,}$ are functions of $g$.

We set $\rho(D-\sqrt \delta)=0$ in (\ref{asymdistr}) and solve this equation order by order in $\ep$ for $r_{0,1,2,}$ and set $\ep=1$ at the end. This yields
\be
r(g)=-2\sqrt {15} \sqrt g+\frac{4A_1}{5}+\frac{135 A_1^2 + 848 A_2}{1500\sqrt{15} \sqrt g}\labell{transline}\ .
\ee
A very reassuring feature of this expression is the $1/\sqrt g$ expansion character. Note, that this has not been put into the equations by hand at any point and is a result of the calculation.

A less reassuring feature of this expression is the fact that after plugged into the expression for $\delta_0$, we obtain and imaginary expression. So we need to look closer at the values of $\delta$. We also need to check which part of the diagram divided by the transition line (\ref{transline}) the asymmetric solutions lives in.

We take the solution $\delta=\delta_0+\ep \delta_1+\ep^2 \delta_2$, plug in $r=r_0+\ep r_1+\ep^2 r_2$ and see that order by order in $\ep$ we obtain real, positive terms. To see which side of the phase transition line the asymmetric one-cut solution exists, we plug $r=r_0+dr_0+\ep (r_1+dr_1)+\ep^2 (r_2+dr_2)$ into the expression for $\delta$ and compute the $\ep$ expansion. We then see that to get a real positive $\delta$, all the steps $dr_{0,1,2}$ need to be negative and thus the existence domain of the asymmetric one-cut solution is under the curve (\ref{transline}).

\begin{figure} [tbp]
\centering % \begin{center}/\end{center} takes some additional vertical space
\includegraphics[width=.75\textwidth]{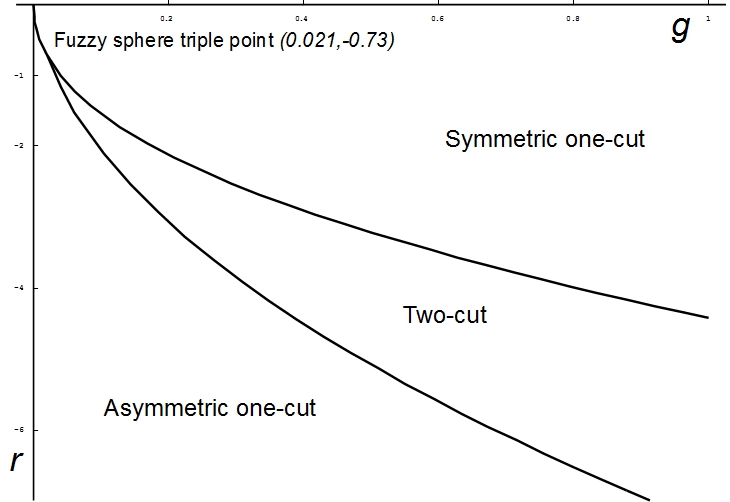}
% "\includegraphics" is very powerful; the graphicx package is already loaded
\caption{Phase diagram of the model. See the text for details.}\labell{fig}
\end{figure}

Finally, let us present the large $N$ phase diagram of the model for the fuzzy sphere in its full glory, figure \ref{fig}. The parameter space is divided into three regions, in which the three possible phases of the theory are realized. The boundary between the asymmetric one-cut region and the two- cut region is given by (\ref{transline}) and the boundary between the two-cut region and the symmetric one-cut region is given by (\ref{12bound}). These two meet at a nontrivial triple point. We do not present the phase diagrams in case of $\CPFn$, as they are qualitatively the same as the phase diagram for the fuzzy sphere.

It is not surprising that repeating the same calculation for the case of the solution living in the negative minimum of the potential yields the very same result, namely the same transition line and the same domain of existence.

There is another way to compute the transition line, which is computationally much more simple, especially when terms of higher order in $\ep$ are considered. It is the same line of attack as we used to compute the symmetric one-cut to two-cut phase transition. We write down the conditions for $\rho(D-\sqrt \delta)=0$, which is an analog of (\ref{trans1}) and two consistency equations for the moments $c_1,c_2$, analogs of (\ref{trans2}). We then solve these pertubratively, without computing general formulas for $\delta$ and $D$ and obtaining formulas valid only at the transition line. With no surprise we find out, that this calculation yields the same results we obtained by the previous approach. However, as this approach does not yield an general expression for $\delta$, it is not usable for showing which part of the parameter space allows for the asymmetric solution.

\paragraph{Location of the triple point for the fuzzy sphere.} We can now locate the triple point. For the fuzzy sphere, by comparing the exact expression (\ref{12boundSF2}) and (\ref{transline}) and numerically computing the intersection of the two curves. We get the triple point
\be
g_c=0.0209 \ , \ r_c=-0.7221\ .\labell{critical1}
\ee

We would like see what happens to this value when higher order contributions to $F(t)$ are considered, mainly as a consistency check that the value does not oscillate wildly and the $1/\sqrt g$ expansion of the phase boundary is well behaved. We should keep in mind that our approximation will miss some terms from the effective action and our results are not going to be valid even up to given order in the perturbative expansion. Repeating the same procedure with terms up to eight order in $t$, i.e. starting with
\be
S_{eff}\approx \half F(c_2-c_1^2)
\ee
and $F(t)$ given by (\ref{generalcpn}) instead of (\ref{themodel}),\footnote{In a general case, we know $F(t)$ only up to seventh order in $t$, we borrow the eight order coefficient $A_8$ from the exact solution (\ref{12boundSF2}).} modifies the two-cut/asymmetric one-cut boundary (\ref{transline}) by new terms that follow the $1/\sqrt g$ expansion character and are given in the appendix \ref{aB}. In the table \ref{tabulka} we present the change of the location of the triple point in the case of the fuzzy sphere.

\begin{table}[tbp]
\centering
\begin{tabular}{|c|ll|}
\hline
order & \multicolumn{1}{c}{$g_c$} & \multicolumn{1}{c|}{$r_c$} \\
\hline 
2 & 0.02091 & -0.7221\\
3 & 0.02127 & -0.7282\\
4 & 0.0215 & -0.7326\\
5 & 0.02171 & -0.7356\\
6 & 0.02185 & -0.7379\\
7 & 0.02195 & -0.7397\\
8 & 0.02204 & -0.7412\\
\hline
\end{tabular}
\caption{Comparison of the location of the triple point in the case of the fuzzy sphere for different perturbative orders.}\labell{tabulka}
\end{table}

We see, that the triple point is not shifted dramatically and thus we can expect the expansion to be well behaved. However we also see that the differences between the values do not vanish very rapidly and it is not clear how far away from the location (\ref{critical1}) the position of the triple point for the full effective action $F(t)$ will be. Either way, we will take the result of our calculation to be
\be
g_c=0.021 \ , \ r_c=-0.73\ ,\labell{critical}
\ee
but will admit that this value does have some uncertainty.

\paragraph{Numerical results.} Let us compare the computed location of the triple point for the fuzzy sphere with the available numerical results from the works \cite{num06,num09,num14}. In our conventions,\footnote{More about the different conventions in the appendix \ref{aA}.} they are as follows
\begin{subequations}
\bea
%g^{\cite{num06}}_c=0.0375\pm0.0125 &,& r^{\cite{num06}}_c=-0.8\pm0.08\ ,\labell{cnum1}\\
%g^{\cite{num09}}_c=0.13\pm0.005 &,& r^{\cite{num09}}_c=-2.3\pm0.2\ ,\labell{cnum2}\\
%g^{\cite{num14}}_c=0.145\pm0.025 &,& r^{\cite{num14}}_c=-2.49\pm0.07\ .\labell{cnum3}
g^{[21]}_c=0.0375\pm0.0125 &,& r^{[21]}_c=-0.8\pm0.08\ ,\labell{cnum1}\\
g^{[19]}_c=0.13\pm0.005 &,& r^{[19]}_c=-2.3\pm0.2\ ,\labell{cnum2}\\
g^{[23]}_c=0.145\pm0.025 &,& r^{[23]}_c=-2.49\pm0.07\ .\labell{cnum3}
\eea
\end{subequations}
Our triple point (\ref{critical}) is at the low end of the numerical precision for the triple point from \cite{num06}, however the two triple points from \cite{num09,num14} are quite different from our value. For a better comparison, we present more complete results of these two papers together with our results in the figure \ref{fignum}.

\begin{figure}[tbp]
\includegraphics[width=.49\textwidth]{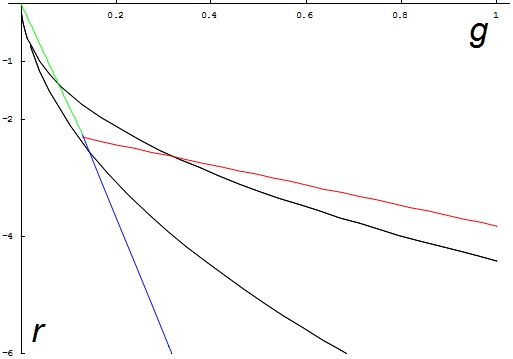}
\includegraphics[width=.49\textwidth]{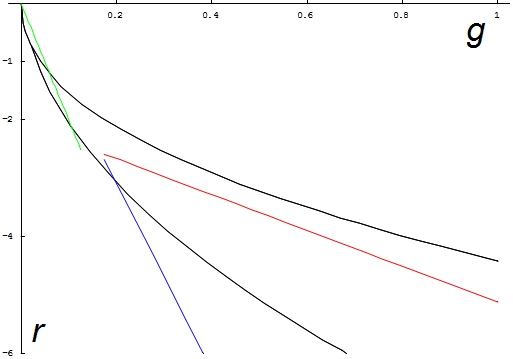}
\caption{The phase transition lines, which are linear fits to the data points, from the numerical work compared with the phase transition lines presented here. The left figure represents the results of \cite{num09} and the right figure represents the results of \cite{num14}. The green lines represent the symmetric one-cut/asymmetric one-cut phase transition, the red lines the symmetric one-cut/two-cut phase transition and the blue lines the two-cut/asymmetric one-cut phase transition.}\labell{fignum}
\end{figure}

In the both cases, the symmetric one-cut/asymmetric one-cut phase transition line (the green line) agrees with the line in our diagram very well. The transition line between the symmetric one-cut and the two-cut phases agrees for larger values of $g$, however the linear fit which is used to extrapolate the transition line to the low values of $g$ (the red line) does not include the dip towards the $g$ axis. This seems to be one of the possible sources of the discrepancy between the values (\ref{critical}) and (\ref{cnum2},\ref{cnum3}).

On the other hand, it is possible that the all orders location of the triple point is higher than our presented value and also it is not clear what the contribution of the terms not included in $F(t)$ is. The difference between the numerical two-cut/asymmetric one-cut phase boundary (the blue line) and our result for larger values of $g$ suggests that the $c_3$ and higher asymmetric terms neglected in the effective actions play an important role in this regime.\footnote{See also a note in the concluding section about the free energy calculation.} On yet another hand, the numerical simulations for the two-cut/asymmetric one-cut phase transition are rather difficult to do as well as the simulations close to the triple point, so there is also some uncertainty of the numerical location.

All this taken into account, we conclude that the value (\ref{critical}) is an reasonable approximation that demonstrates the validity of our method, but probably underestimates the true value and we expect the neglected contributions to shift this position to higher values.

\paragraph{Location of the triple point for a general $\CPFn$.} For the case of the $\CPFn$, we do not have an exact formula for the one-cut/two-cut boundary, so to locate the triple point we can use only the approximation obtained in section \ref{32}. Repeating essentially the same procedure as for the fuzzy sphere triple point, we obtain an expression for the triple point for a general $\CPFn$. Since the expression is only approximate and is very long and cumbersome, we do not present it. We only summarize the values for the triple point for the first few $\CPFn$'s in the table \ref{tabulka2}.

\begin{table}[tbp]
\centering
\begin{tabular}{|c|ll|}
\hline
n & \multicolumn{1}{c}{$g_c$} & \multicolumn{1}{c|}{$r_c$} \\
\hline
1 & 0.02160 & -0.7369\\
2 & 0.1714 & -2.048\\
3 & 0.9320 & -5.333\\
4 & 2.890 & -9.784\\
5 & 6.713 & -15.20\\
\hline
\end{tabular}
\caption{Numerical values for the $\CPFn$ triple points. The value for $n=1$, i.e. the fuzzy sphere, was obtained using the approximate phase boundary and not using the exact expression.}\labell{tabulka2}
\end{table}

\section{Conclusions}
We have studied the phase structure of the fuzzy $\CPFn$ scalar $\phi^4$ theory. After the integration of the kinetic part of the action the field theory became a hermitian matrix model. We have analyzed the eigenvalue distribution of this model and have shown, that this model admits an asymmetric one-cut distribution, apart from the standard one-cut and two-cut distributions. We have computed the boundaries between the domains of existence of these phases in the parameter space of the theory and identified a triple point, i.e. the point where the boundaries meet. We have computed the location of this triple point and it is not too far away from the location obtained in previous numerical simulations.

Several approximations have been made in the process of our calculation. We have assumed that the eigenvalue distribution is not too far from the semicircle, that it is not very asymmetric and we have considered only first few terms in the expansion of the kinetic term effective action. There is clearly plenty of space for improvement here and we expect that when the neglected terms are taken into account, the triple point will be shifted away from the origin of the parameter space and closer to the numerical location. Also, at the moment there is no systematic approach to the neglected terms and it would be interesting to see whether one can find a procedure to fully determine them.

We have considered a special type of an asymmetric solution. A more generic asymmetric solution is given by a two-cut solution with both wells of the true potential filled with an unequal number of eigenvalues. The phase transition occurs when one of the wells becomes empty. To analyze this process completely, one has to compute the free energy of the asymmetric two-cut phase and determine when the complete filling of one of the wells is energetically preferred. This could shift our phase transition line into the region where the asymmetric one-cut solution is allowed, but we do not expect this to be significant for small values of $g$ where the triple point is located.

The presence of the non-uniform order phase in the phase diagram of the field theory, even in the large $N$ limit, is a manifestation of UV/IR-mixing. Modifications of the $\phi^4$ theory with no UV/IR-mixing are available \cite{uvir,gw,nouvir2} and it should be possible to analyze the phase diagram of these with our method. The modification is expected to push the location of the triple point to higher values, ultimately removing it from the diagram, together with the domain of existence of the non-uniform order solution.

Recent numerical study \cite{num14panero2} shows that the double cut phase and the triple point survive also the infinite volume limit, in which the fuzzy sphere becomes the noncommutative plane. It would be interesting to study such limit of the phase diagram and obtain this feature analytically. Moreover, it should be possible to do the same for the higher dimensional cases. Apart from the infinite volume limits of $\CPFn$, there are several other fuzzy field theories which are all more or less straightforwardly suited for the approach presented here, namely a formulation of the field theory on the fuzzy $4$-sphere \cite{s4}, numerical and perturbative calculations for the case of $\mathbb R\times S_F^2$ \cite{rsfnum,rsfper} and a formulation of the field theory in a three dimensional noncommutative space \cite{r3nc}. It would be interesting to see, if we could reconstruct previous results and what kind of new results could be obtained.

\acknowledgments
This work was supported by the \emph{Alumni FMFI} foundation as a part of the \emph{N\'{a}vrat teoretikov} project.

%\paragraph{Note added.} This is also a good position for notes added after the paper has been written.

\appendix
\section{Conventions dictionary}\labell{aA}
In the paper, we refer heavily to three previous works \cite{samann,num09,num14}, which all use a little different conventions from ours. We thus present a short summary of the differences. We denote the quantities as in \cite{samann} by a bar and quantities as in \cite{num09,num14} by a hat.

In \cite{samann}, the action has been
\bea
\bar S&=&\beta\,\trl{\bar M [\bar L_i,[\bar L_i,\bar M]]+\bar r \bar M^2+\bar g \bar M^4}\ ,\\
&& [\bar L_i,[\bar L_i,T^l_m]]=2l(l+n)T^l_m\ ,
\eea
which yields
\be
\bar M=\frac{M}{\sqrt{4\beta}} \ , \ \bar r = 2r \ , \ \bar g=16\beta g\ .
\ee
For the scalings, we obtain
\be \theta_x=\frac{\bar \theta_\lambda+\half \bar \theta_\beta}{n} \ , \ \theta_r=\frac{\bar \theta r}{n} \ , \ \theta_g=\frac{\bar \theta_g - \bar \theta_\beta}{n}\ .\ee

In the both numerical works \cite{num09,num14}, the action used has been
\bea
\hat S&=&\trl{ \hat M [ \hat L_i ,[\hat L_i ,\hat M ]]+b \hat M^2+c \hat M^4}\ ,\\
&& [\tilde L_i,[\tilde L_i,T^l_m]]=l(l+1)T^l_m\ ,
\eea
which yields
\be
\hat M=\frac{M}{\sqrt{2}} \ , \ b = r \ , \ c=4 g\ .
\ee

\section{Some explicit formulas}\labell{aB}
For completeness and future reference, let us present full results of our perturbative calculations that were omitted in the text.

For a general $\CPn$, the effective action is
\be\labell{generalcpn}
F(t)=A_1 t + A_2 t^2+ A_3 t^3+ A_4 t^4+ A_5 t^5+ A_6 t^6+ A_7 t^7,
\ee
where
\begin{subequations}
\bal
A_1&=\frac{n(n!)^{2/n}}{n+1}\ , \ A_2=-\frac{n(n!)^{4/n}}{2(n+1)^2(2+n)}
\ , \ 
A_3=-\frac{2(n-1)n(n!)^{6/n}}{3(n+1)^3(6+5n+n^2)}
\\
A_4&=-\frac{n(12-24n+n^2+7n^3)(n!)^{8/n}}{4(n+1)^4(n+2)^2(12+7n+n^2)}
\\
A_5&=-\frac{2n(-24+92n-77n^2-8n^3+17n^4)(n!)^{10/n}}{5(n+1)^5(n+2)^2(60+47n+12n^2+n^3)}
\\
A_6&=-\frac{n\lr{\begin{array}{lr}720-3660n+3854n^2+1524 n^3-2429 n^4-439 n^5+\\423 n^6+10n^7\end{array}}(n!)^{12/n}}{3(n+1)^6(n+2)^3(n+3)^2(120+74n+15 n^2+n^3)}
\\
A_7&=-\frac{12n\lr{\begin{array}{lr}-720+5304 n - 10960 n^2+5426 n^3+4521 n^4 - 3282 n^5-\\884 n^6+472 n^7+128n^8\end{array}}(n!)^{14/n}}{7(n+1)^7(n+2)^3(n+2)^2(840+638n+179 n^2+22 n^3+n^4)}
\end{align}
\end{subequations}

The following are the explicit formulas for the approximations to the effective action using the formula (\ref{ht}). The effective action is the logarithm of the given expression, $F(t)=\log\lr{\ldots}$
\begin{subequations}
\begin{align}
n=2 \ &,\ 
\frac{\lr{\begin{array}{lr}-11789184402450 + 93978027360 t + 5998294160460 t^2\\ + 2932050372000 t^3 + 497642752529t^4\end{array}}}{30(-392972813415+
527096352132t - 197206475949 t^2 + 39981049252 t^3)}\ ,\\
n=3 \ &,\ 
\frac{\lr{\begin{array}{lr}-824984552000 + 272764119600 6^{2/3} t + 1454433162840\,6^{1/3} t^2\\ + 1945660952760 t^3 + 167744125503\, 6^{2/3}
t^4\end{array}}}{560(-1473186700+ 1591968810\, 6^{2/3} t - 2246388516\, 6^{1/3} t^2 + 1411751691\, t^3)}\ ,\\
n=4 \ &,\   
\frac{\lr{\begin{array}{lr}148500768040625 + 25279264566220000\sqrt6 t + 150340268216376000 t^2\\ + 60165687487957760\sqrt6 t^3 + 60941670673759896 t^4\end{array}}}{5\lr{\begin{array}{lr}29700153608125 + 5008332667471000\sqrt6 t\\ - 18230533095002200 t^2 + 4236389020569472\sqrt6 t^3\end{array}}}\ ,\\
n=5 \ &,\ 
\frac{\lr{\begin{array}{lr}
159606423231255\, 2^{4/5} 15^{3/5} + 9098272445481450\, t\\ + 6433230719486025\, 2^{1/5} 15^{2/5} t^2
 + 2256395876247656\, 2^{2/5} 15^{4/5} t^3\\ + 5382149790792650\, 2^{3/5} 15^{1/5} t^4\end{array}}}{\lr{\begin{array}{lr}
159606423231255\, 2^{4/5} 15^{3/5} + 1117951283918700 t -\\
1890281408310225\, 2^{1/5} 15^{2/5} t^2 + 562828122041281\, 2^{2/5} 15^{4/5} t^3\end{array}}}\ .
\end{align}
\end{subequations}

Finally, we present the result of the perturbative calculation for the two-cut/asymmetric one-cut phase boundary up to the eight order in the effective action
\begin{align}
r(g)&=-2\sqrt 15 g+\frac{4 A_1}{5}+\lr{\frac{135 A_1^2+848 A_2}{1500 \sqrt 15}}\frac{1}{\sqrt g}
\\&
+\frac{1}{g}\lr{\frac{2025 A_1^3+11 236 A_3+10170 A_1 A_2}{562500}}
\no&
+\frac{1}{(\sqrt g)^3}\frac{
\lr{\begin{array}{lr}5740875 A_1^4 + 32464800 A_1^2 A_2 + 19805040 A_2^2 \\
+29707560 A_1 A_3 + 19056256 A_4\end{array}}}{2025000000 \sqrt 15}
\no&
+\frac{1}{g^2}\frac{
\lr{\begin{array}{lr}19683000 A_1^5 + 108172665 A_1^2 A_3 + 70685676 A_1 A_4 + 31561924 A_5\\
 + 126372150 A_1^3 A_2 + 106028514 A_3 A_2 + 144230220 A_1 A_2^2
\end{array}}}{113906250000}
\no&
+\frac{1}{(\sqrt g)^5}\frac{
\lr{\begin{array}{lr}237487696875 A_1^6 + 1709468550000 A_1^4 A_2 + 2801052738000 A_1^2 A_2^2 \\
+  609290035200 A_2^3 + 1400526369000 A_1^3 A_3 + 2741805158400 A_1 A_2 A_3 \\
+ 423995740920 A_3^2 + 913935052800 A_1^2 A_4 + 753770206080 A_2 A_4 \\
+ 471106378800 A_1 A_5 + 26764511552 A_6
\end{array}}}{1366875000000000 \sqrt 15}
\no&
+\frac{1}{g^3}\frac{
\lr{\begin{array}{lr}
2391484500000 A_1^7 + 15074840643750 A_1^4 A_3 + 9480160743525 A_1 A_3^2  \\
+ 9749822053500 A_1^3 A_4 + 4512250464660 A_3 A_4 + 5266755968625 A_1^2 A_5  \\
 +376020872055 A_1 A_6 + 620602111612 A_7 + 19049945512500 A_1^5 A_2 \\
+ 43874199240750 A_1^2 A_3 A_2 + 16853619099600 A_1 A_4 A_2 \\+  3760208720550 A_5 A_2
+40199575050000 A_1^3 A_2^2 \\+  12640214324700 A_3 A_2^2 + 19499644107000 A_1 A_2^3
\end{array}}}{192216796875000000}
\no&
+\frac{1}{(\sqrt g)^7}\frac{
\lr{\begin{array}{lr}
51778255111171875 A_1^8 + 451444780593000000 A1^6 A_2  \\
 +1156207911142500000 A_1^4 A_2^2 + 886545939600000000 A_1^2 A_2^3  \\
 +97520988528864000 A_2^4 + 346862373342750000 A_1^5 A_3  \\
 +1329818909400000000 A_1^3 A_2 A_3 + 877688896759776000 A_1 A_2^2 A_3  \\
 +329133336284916000 A_1^2 A_3^2 + 170459137959667200 A_2 A_3^2  \\
 +221636484900000000 A_1^4 A_4 + 585125931173184000 A_1^2 A_2 A_4  \\
 +151519233741926400 A_2^2 A_4 + 227278850612889600 A_1 A_3 A_4 \\
 +23876085465308160 A_4^2 + 121901235661080000 A_1^3 A_5  \\
 +189399042177408000 A_1 A_2 A_5 + 44767660247452800 A_3 A_5  \\
 +9469952108870400 A_1^2 A_6 + 5969021366327040 A_2 A_6  \\
 +20891574782144640 A_1 A_7 + 4811616828772352 A_8\end{array}}}{3690562500000000000000\sqrt 15}\ .\nonumber
\end{align}

\end{document}